# Crossover from individual to collective magnetism in dense nanoparticle systems: local anisotropy *vs* dipolar interactions

*Elena H. Sánchez\*, Marianna Vasilakaki, Su Seong Lee, Peter S. Normile, Mikael S. Andersson, Roland Mathieu, Alberto López-Ortega, Benoit P. Pichon, Davide Peddis, Chris Binns, Per Nordblad, Kalliopi Trohidou, Josep Nogués\* and José A. De Toro\**

Elena H. Sánchez, Peter S. Normile, Chris Binns, Alberto López-Ortega and José A. De Toro
Instituto Regional de Investigación Científica Aplicada (IRICA) and Departamento de Física Aplicada, Universidad de Castilla-La Mancha, 13071 Ciudad Real, Spain

Marianna Vasilakaki and Kalliopi Trohidou
Institute of Nanoscience and Nanotechnology NCSR "Demokritos", 153 10 Agia Paraskevi, Greece

Su Seong Lee
 Institute of Materials Research and Engineering, 31 Biopolis Way, #09-01, The Nanos, Singapore 138669

Mikael S. Andersson
Department of Chemistry, Ångström Laboratory, Uppsala University, 75121 Uppsala, Sweden
Department of Chemistry and Chemical Engineering, Chalmers University of Technology, SE-412 96 Göteborg, Sweden

Roland Mathieu and Per Nordblad
Department of Materials Science and Engineering, Uppsala University, Box 35, 751 03, Uppsala, Sweden

Alberto López-Ortega
Departamento de Ciencias, Universidad Pública de Navarra, 31006 Pamplona, Spain
Institute for Advanced Materials and Mathematics (INAMAT2), Universidad Pública de Navarra, 31006 Pamplona, Spain

Benoit P. Pichon
Université de Strasbourg, CNRS, Institut de Physique et Chimie des Matériaux de Strasbourg, UMR 7504, Strasbourg F-67000, France
Institut Universitaire de France, 75231 Paris Cedex 05, France

Davide Peddis
Università degli Studi di Genova, Dipartimento di Chimica e Chimica Industriale, Via Dodecaneso 31, 1-16146 Genova, Italy
Istituto di Struttura della Materia-CNR, 00015 Monterotondo Scalo (RM), Italy

Josep Nogués
Catalan Institute of Nanoscience and Nanotechnology (ICN2), CSIC and BIST, Campus UAB, Bellaterra, 08193 Barcelona, Spain
ICREA, Pg. Lluís Companys 23, 08010 Barcelona, Spain

E-mails: elena.hdez.schez@gmail.com, josep.nogues@icn2.cat, JoseAngel.Toro@uclm.es

**Abstract**

Dense systems of magnetic nanoparticles may exhibit dipolar collective behavior. However, two fundamental questions remain unsolved: (i) whether the transition temperature may be affected by the particles anisotropy or it is essentially determined by the intensity of the interparticle dipolar interactions, and (ii) what is the minimum ratio of dipole-dipole interaction ($E_{dd}$) to nanoparticle anisotropy ($K_{ef}V$, anisotropy·volume) energies necessary to crossover from individual to collective behavior. We have studied a series of particle assemblies where dipolar interactions are similarly intense, but the nanoparticle anisotropy widely varies across the series. This parameter has been tuned through different degrees of cobalt-doping in maghemite nanoparticles, resulting in a variation of nearly an order of magnitude. All the (essentially bare) particle compacts display collective behavior, except the one made with the highest anisotropy particles, which presents "marginal" features. Thus, a threshold of $K_{ef}V/E_{dd} \approx 130$ to suppress collective behavior is derived, in good agreement with Monte Carlo simulations. This translates into a crossover value of $\approx 1.7$ for the easily accessible parameter $T_{MAX}(interacting)/T_{MAX}(non-interacting)$ (temperature ratio of the maximum in the temperature dependent zero-field-cooled magnetization of interacting and dilute particle systems), which has been successfully tested against the literature to predict the individual-like/collective behavior of any given interacting particle assembly comprising relatively uniform particles.

# 1. Introduction

Dense assemblies of magnetic nanoparticles continue to attract high interest, both from the technological[1–4] and basic science (*e.g.*, superspin-glass dynamics, collective *vs* individual behavior)[5–11] viewpoints. The magnetic properties of the assembly depend markedly on the concentration. While dilute systems show magnetic properties similar to those of the individual particles, sufficiently concentrated assemblies may exhibit collective behavior, with properties (*e.g.*, transition temperature or coercivity) distinctly different from those of their constituent particles. Dipolar collective behavior in magnetic nanoparticle systems can be beneficial in applications such as magnetic resonance imaging and hyperthermia,[12,13] and detrimental to the performance of others (such as magnetic storage, permanent magnets or magnetoresistive sensors).[14–16] In addition, in the recently discovered "liquid permanent magnets" dipolar interactions may play a role in the magnetic stabilization of the particles jammed at the interface of immiscible liquids.[17] Yet, the fundamental question as to the relative importance of the local anisotropy and the interparticle interactions in the blocking/freezing temperature and other magnetic properties of nanoparticle assemblies remains an open question, despite the relevance of this issue in many applications and the fundamental incentive provided by the contrasting theoretical descriptions of the effect. Namely, Mørup proposed in his seminal phase diagram that the freezing temperature (peak temperature in zero-field-cooled magnetization curves of systems exhibiting collective behavior) is simply proportional to the dipole-dipole interaction strength,[18] whereas the anisotropy energy barrier of the individual nanoparticles is explicitly considered in the Vogel-Fulcher approach, also used to describe interacting nanoparticles.[19–21]

It is important to emphasize that while the behavior of particle systems with weak dipolar interactions can be simply accounted for by a modified Néel-Brown model, where the anisotropy of the particles still plays the main role in determining the blocking temperature,[22] dense random systems of dipolarly-interacting nanoparticles have far more complex characteristics. In particular, they have been extensively reported to exhibit a collective (superspin glass) transition similar to that in conventional (atomic) spin glass freezing.[6,22–26] However, a fundamental difference between superspin- and classical spin-glass systems is the existence in the former of random (nanoparticle) anisotropy barriers ($K_{ef}V$) yielding strongly temperature-dependent local relaxation times. Their effect on the overall magnetic properties of dense assemblies has rarely been studied experimentally.[11] The abundant experimental literature on strongly interacting particle systems has delved into the effects of dipolar interactions *by varying the concentration of the magnetic nanoparticles*, first in frozen ferrofluids[21,27–29] and nanogranular alloys in thin film or powder form,[30–32] and more recently by controlling the thickness of non-magnetic spacers (such as silica or dendrimer coatings)[20,22,24,33–35] or the degree of powder compaction.[36]

Here, our experimental design takes on the opposite approach; namely, the packing fraction has been fixed (~ 60%, close to the theoretical maximum of random-close-packing of hard spheres,[37] in order to procure the strongest possible dipolar interactions), and the particle anisotropy systematically varied across an order of magnitude (by using maghemite nanoparticles doped with different amounts of Co).[38,39] Consequently, this shifts the focus from interparticle interactions to the importance of nanoparticle magnetic anisotropy. This strategy has allowed us to tackle the unsolved issue of the role of the *local* anisotropy barrier on the *collective* characteristics of dense assemblies of nanoparticles. The results, backed by Monte Carlo simulations and a literature review, show that sufficiently large local anisotropies will suppress the collective behavior of the assembly. In addition, they offer an estimate of the ratio of the relevant energies (anisotropy barrier and dipole-dipole interaction) that yields a crossover from individual to collective dynamics.

## 2. Results and Discussion

### 2.1. Preparation of dense nanoparticle systems

Four batches of highly monodisperse nanoparticles were synthesized through thermal decomposition using oleic acid (OA) as surfactant:[8,40] pure maghemite ($\gamma$-Fe$_2$O$_3$) nanoparticles with a mean diameter of 6.9 ± 0.6 nm and three types of cobalt-doped maghemite particles with similar sizes (6.7 ± 0.6 nm) and different $f_{Co}$ = Co/(Co+Fe) ratio, namely $f_{Co}$ = 0.11, 0.19 and 0.23. **Figure 1** shows the transmission electron microscopy (TEM) images and particle size histograms, confirming the almost identical nanoparticle diameter independently of the Co-doping. The narrow size distribution shown is representative of all the particles, with a polydispersity lower than 4%. Electron energy-loss spectroscopy (EELS) mapping analysis was performed to discard cation segregation during the synthesis. The images (see Figure S1) reveal a homogeneous distribution of the cobalt and iron ions across the whole nanoparticle, ruling out phase separation.

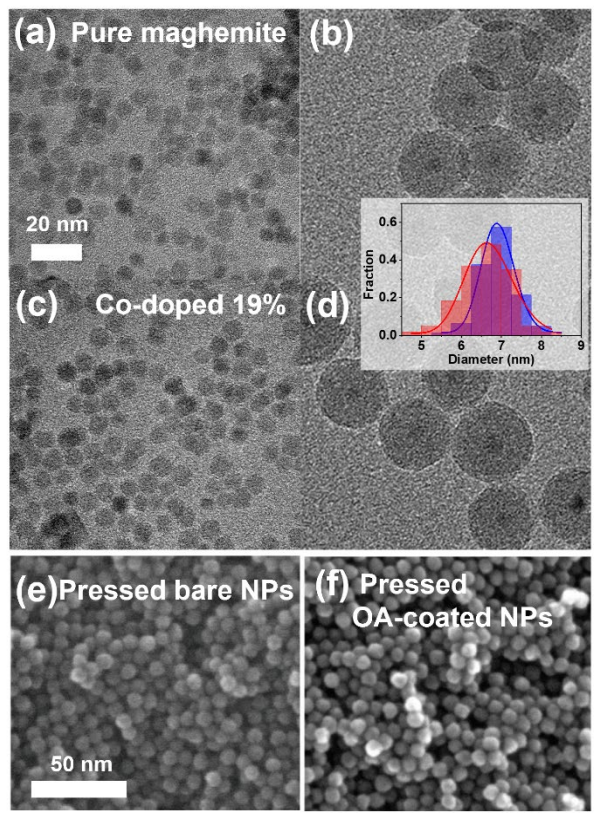

**Figure 1.** Transmission electron microscopy (TEM) images of the pure maghemite (a, b) and Co-doped 19% maghemite particles (c, d). The silica-coated particles are shown on the right side, and the OA-coated particles on the left side (scale bar = 20 nm). The (mostly overlapping) size distributions of both samples (pure -blue- and doped -red-) are shown in the inset. Typical high-resolution scanning electrons microscopy (HRSEM) images of the surface of discs prepared with bare (e) and OA-coated (f) nanoparticles.

After the chemical synthesis, a fraction of the particles was washed with acetone several times to remove the oleic acid surfactant bound to the particles, producing essentially bare particles. Figure S2 shows the thermogravimetric curves measured to quantify the content of oleic acid before ($\approx$ 20%, very close to the expected value for a monolayer of oleic acid) and after ($\approx$ 5%) the described washing.

Another fraction of each type (doping) of particles was coated with thick silica ($SiO_2$) shells of thickness around 17 nm [Figure 1(b) and (d)] in order to produce reference magnetically dilute systems with negligible interparticle interactions.[22,40–42] Therefore, three series of samples were prepared: silica-coated, OA-coated and bare nanoparticles, with different degrees of Co-doping. All powders were subsequently compacted under a uniaxial pressure of $\approx$ 1 GPa to make dense discs [see Figure 1(e,f)], resulting in the "magnetic" concentration values of C = 0.4% (silica-coated), 50% (OA-coated), and 60% (bare nanoparticles), respectively.[43] These pellets represent ideal non-interacting (silica-coated nanoparticles), moderately interacting (OA-coated nanoparticles) and strongly interacting (bare nanoparticles) systems.

## 2.2. Magnetic characterization

**Figure 2** depicts the zero-field-cooled (ZFC) magnetization curves for the three series of samples described above: (a) silica-coated nanoparticles, (b) OA-coated nanoparticles and (c) bare nanoparticles. All the samples exhibit a ZFC maximum ($T_{MAX}$) that increases, in all three series, with the content of Co [see also Figure 3(a), and Table 1 for numerical values]. The *silica-coated* particle assemblies [Figure 2(a)] provide the single particle behavior, as the thick diamagnetic shell makes magnetostatic interactions negligible.[8,22] In these systems, the relaxation time of the nanoparticle macrospin is simply governed by the ratio of the thermal ($k_B T$) to the anisotropy barrier ($K_{ef}V$) energies, where $K_{ef}$ is the effective uniaxial anisotropy constant of the particle, as described by the Néel-Brown model $\tau = \tau_0 \exp\left(K_{ef}V/k_B T\right)$. When this relaxation time is smaller (greater) than the characteristic time of the measurement technique ($\tau_m$), the magnetic response of the assembly is described as "superparamagnetic" ("blocked"). The crossover temperature between the two regimes (for which $\tau = \tau_m$) is known as the blocking temperature ($T_B$) and is directly proportional to the product of the nanoparticle volume and its effective anisotropy constant ($K_{ef}V$). Therefore, given the constant volume across the *silica-coated* series the increase in $T_{MAX}$ (often taken as $T_B$ in these kind of dilute systems) across the series clearly indicates an anisotropy ($K_{ef}$) enhancement due to the introduction of cobalt cations in the spinel structure.[38] Note that the same trends are observed when using the more accurate mean blocking temperature $T_{Bm}$ (defined as the peak temperature in the $-d(M_{FC}-M_{ZFC})/dT$ curve,[44–46] see Figure S3). The effective anisotropy constant can be readily calculated from the Néel-Brown equation using the $T_{Bm}$ values, the measurement time $\tau_m$ = 100 s (typical for dc SQUID magnetometry) and the attempt time $\tau_0$ = 10$^{-13}$ s, yielding $K_{ef}$ ≈ 32, 228, 282 and 317 kJ m$^{-3}$ for the pure maghemite, 11%, 19% and 23% Co-doped particles, respectively. Note that this attempt time value has been previously found to systematically fit the data of similar isolated particles better than the customary[22,23,26] $\tau_0 = 10^{-10}$ s (in any case, this choice introduces a factor of only 1.25 in $K_{ef}$), although faster attempt times have also been reported in both metallic[47] and spinel cobalt ferrite particles.[48] Comparable $K_{ef}$ values have been reported in the literature, *e.g.*, 140 and 250 kJ m$^{-3}$ (using $\tau_0$ = 10$^{-10}$ s) for Co-doped iron oxide particles with similar size (5 and 6 nm, respectively) and cobalt content (15 and 19%, respectively).[38,49]

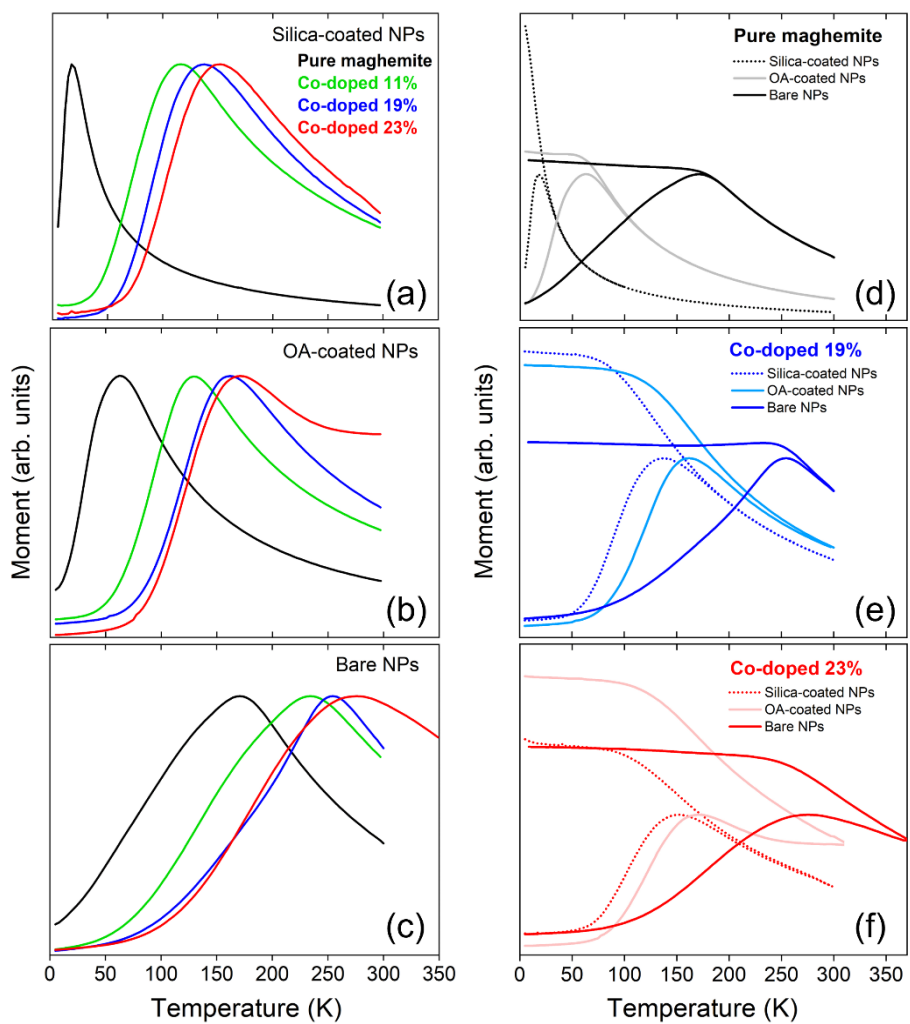

**Figure 2.** The left column shows the ZFC magnetization curves normalized by $M_{ZFC}(T_{MAX})$ measured in an applied field of $\mu_0 H = 0.5$ mT in pressed discs of (a) nanoparticles coated with a thick silica shell, (b) OA-coated nanoparticles, and (c) bare nanoparticles. The right column shows the ZFC and FC magnetization curves (measured in the same conditions) of (d) pure maghemite, (e) 19% and (f) 23% Co-doping particles with different coatings: silica (dashed lines), oleic acid (solid light-colored lines) and bare (solid dark-colored lines).

In the series of pressed OA-coated nanoparticles [Figure 2(b)], the increase of $T_{MAX}$ with respect the corresponding isolated nanoparticles is stronger for the pure maghemite particles ($\Delta T = 45$ K) than for the three Co-doped particle systems ($\Delta T \approx 20$ K). Interestingly, this offers a first indication that the $T_{MAX}$ shift is not determined solely by interparticle interactions, but it also depends on the nanoparticle anisotropy barrier. The OA-coated Co-doped samples are identified as "weakly/moderately interacting" systems with dynamics still described by a modified Arrhenius law,[50]

$$\tau = \tau_0 e^{\frac{K_{ef}V + \Delta}{k_B T}}, \qquad (1)$$

where the dipolar interactions simply enhance by an amount $\Delta$ the single-particle anisotropy. This "perturbation" description is suitable only for relatively small values of the $\Delta/K_{ef}V$ ratio, which explains why the lowest anisotropy particles (pure maghemite) of the OA-coated series showed collective, rather than modified single particle, behavior.[22] In a first, qualitative, approach, the collective behavior of this sample can be identified from the flat shape of the FC magnetization curve below $T_{MAX}$ *with a magnetization close to $M_{ZFC}(T_{MAX})$* [see grey line in Figure 2(d)]. Note that a flat FC curve at low temperature is not indicative on its own of collective behavior; in fact, it is customarily observed in systems of isolated particles with a narrow size distribution and high enough $T_{MAX}$, as is the case of our silica-coated Co-doped nanoparticles. On the other hand, in the three pellets comprising high-anisotropy OA-coated Co-doped nanoparticles, similarly strong dipolar interactions (in fact considerably stronger than those in traditional "dense" frozen ferrofluids or granular solids)[5,25,30,51–53] were not enough to produce collective behavior [see, *e.g.*, the FC curves (light-colored) of the 19% and 23% doping samples in Figure 2(e) and (f), comparable to the FC curves of the corresponding isolated nanoparticles (dashed lines)].

**Table 1**. Peak temperatures ($T_{MAX}$) measured in the ZFC magnetization curves of the 12 samples (pressed discs) studied in this work. The error is less than 1 K in all cases. The Co-doping atomic ratio is defined as $f_{Co}$ = Co/(Co + Fe). The estimated particle concentration (or packing fraction, $C$) for each type of particle coating is also indicated.[40,43]

| | $T_{MAX}$ (K) | | |
|---|---|---|---|
| $f_{Co}$ | *Silica-coated nanoparticles* ($C \approx 0.4\%$) | *OA-coated nanoparticles* ($C \approx 50\%$) | *Bare nanoparticles* ($C \approx 60\%$) |
| 0 | 17 | 62 | 170 |
| 0.11 | 114 | 129 | 234 |
| 0.19 | 138 | 161 | 255 |
| 0.23 | 153 | 171 | 275 |

Remarkably, for the assemblies prepared from the uncoated (bare) particles, all samples exhibited flat FC curves at low temperatures indicating collective behavior, although clearly modulated by the varying nanoparticle anisotropy, except possibly the pellets made with the 23%-doped nanoparticles [see FC curves in Figure 2(e-f)]. To better illustrate the above FC-ZFC irreversibility qualitative argument, Figure S4 plots a parameter expressing the rise of the FC plateau above the ZFC peak, $FC_{rise} = [M_{plateau} - M_{ZFC}(T_{MAX})]/M_{ZFC}(T_{MAX})$ for the different particle systems.

Note that in the OA-coated series, the pressed oleic acid introduces a uniform separation of about 1.3 nm between nanoparticles (estimated from the "magnetic" packing fraction),[43] ruling out the possibility of interparticle exchange coupling. Importantly, even in the *bare* nanoparticle series direct exchange coupling is also expected to play only a marginal role.[22] Firstly, it is rather counterintuitive that the delicate indirect exchange at play in ferrites can propagate between metal ions belonging to different particles, unless there exists an exceptional crystalline coherence between aligned particles.[54,55] In addition, that the dominating type of interaction in the *bare* series is dipolar coupling is hinted by the very similar exchange bias fields ($H_E$) measured at 5 K in the discs made of Co-doped nanoparticles with or without coating (see Figure S5 and the corresponding discussion). In short, $H_E$ in these samples is ascribed to the presence of surface spin disorder,[56–62] which would be affected by interparticle exchange interactions should there be present in the *bare* nanoparticle systems. The residual oleic acid bound to the nanoparticles will further hamper interparticle exchange. For a more detailed discussion, see the Supporting Information in Ref. [8].

As displayed in Figure 2(c), the strong interactions in the *bare* particle compacts enhance the ZFC peak temperatures well beyond the values measured in the *OA-coated* series. Moreover, the flat FC curves, combined with the more quantitative features that will be presented later (*i.e.*, ZFC memory effect and critical slowing down of the relaxation in SI), indicate a crossover from single-particle to collective (superspin glass like) dynamics with increasing packing fraction of the magnetic Particles. Therefore, in this series, the measured peak temperature ($T_{MAX}$) reflects the collective freezing of the system upon cooling.[23,26] Yet, despite this collective character (see discussion below), $T_{MAX}$ is still not only controlled by the dipolar interactions, but it is also largely affected by the local anisotropy (nanoparticle energy barrier). This result is at odds with the conceptual phase diagram proposed by Mørup for dipolarly interacting particle systems, where the collective freezing temperature is determined exclusively by the strength of the dipolar interactions.[18] In fact, **Figure 3(a)** shows that $T_{MAX}$ increases with nanoparticle anisotropy (*i.e.*, $f_{Co}$) despite the concomitant *reduction* in dipolar interaction strength (green arrow in the figure) stemming from the small loss in saturation magnetization upon Co-doping (see Figure S6 and the corresponding discussion).[63,64] This small variation of dipolar interactions across the $f_{Co}$ series is an undesired effect of our Co-doping strategy, yet it lends support to the argument.

Figure 3(a), therefore, shows that $T_{MAX}$ in any dipolar interacting nanoparticle system is determined by additive contributions from the dipolar interactions (characterized by the parameter $T_{dd} = \mu_0 M_S^2 V^2 / k_B 4\pi r^3$, where $r$ is the mean distance between particles) *as well as* from the nanoparticle anisotropy energy barrier ($K_{ef}V$, here estimated from $T_{Bm}$, measured in the silica-coated nanoparticles). However, modified single-particle dynamics (**Equation 1**), where the relaxation time ($\tau$) still diverges only at T = 0 K, does not hold for more strongly interacting nanoparticle systems where $\tau$ is found to diverge at a finite (glass) temperature, signaling a phase transition to a superspin glass state. A phenomenological expression satisfying both requisites [namely, (i) $T_{MAX}$ determined by both $T_{dd}$ and $T_{Bm}$, and (ii) critical divergence of the relaxation time at non-zero temperature] is the well-known Vogel-Fulcher law:[19]

$$\tau = \tau_0 \, e^{\frac{K_{ef}V}{k_B(T - T_{int})}} \quad (2)$$

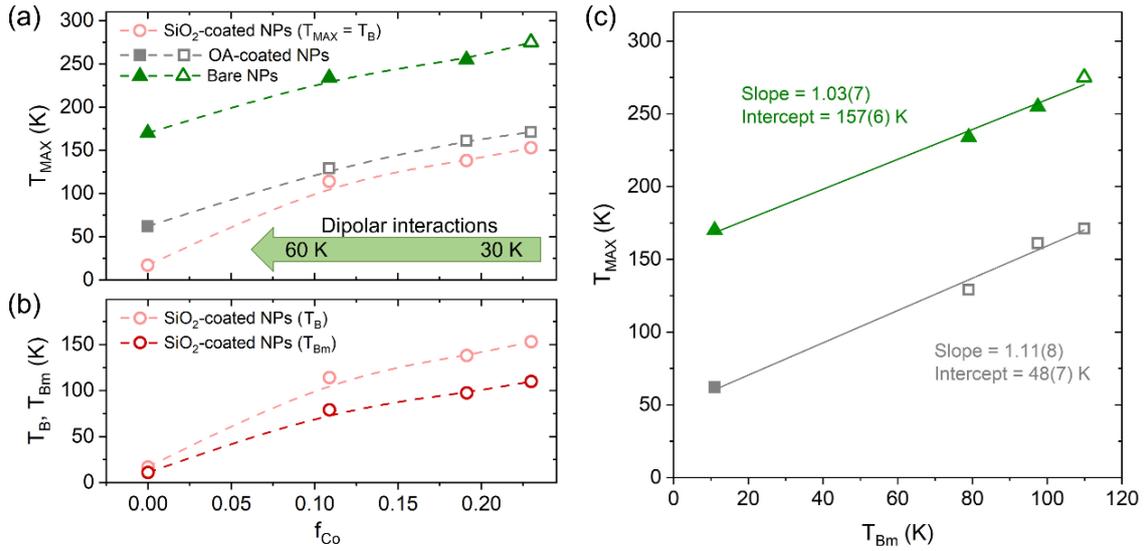

**Figure 3.** (a) $T_{MAX}$ (obtained from the peak of the ZFC curve) as a function of the cobalt-doping fraction ($f_{Co}$) for pressed discs comprising bare, OA-coated and silica-coated identical particles. For the reference (dilute) system (nanoparticles coated with a thick silica shell), $T_{MAX}$ is defined as the blocking temperature $T_B$. The mean blocking temperature ($T_{Bm}$) derived from the FC-ZFC difference in the dilute systems (see text) is also plotted in panel (b). The dashed lines are guides to the eye and the green arrow indicates the decreasing dipolar interaction strength ($T_{dd}$) with increasing Co fraction (the range of the variation is estimated for the *bare* nanoparticle series). (c) $T_{MAX}$ (for the interacting systems) plotted as a function of $T_{Bm}$ measured in the isolated (silica-coated) nanoparticles. Hollow and solid symbols correspond to systems with individual and collective behavior, respectively. The solid lines are linear fits.

For $\tau = \tau_m$, $T = T_{MAX}$, and solving for the latter:

$$T_{MAX} = \left(1/\ln\left(\frac{\tau_m}{\tau_0}\right)\right) \frac{K_{ef}V}{k_B} + T_{int} = T_{Bm} + b_s T_{dd} \quad (3)$$

where $T_{Bm}$ is the mean blocking temperature measured by magnetometry in isolated particles ($K_{ef}V$ is taken here as the mean anisotropy barrier of the nanoparticles), and $b_s$ is a proportionality constant relating the Vogel-Fulcher divergence temperature $T_{int}$, customarily ascribed to interparticle interactions,[19–21] and the dipole-dipole interaction temperature $T_{dd}$ defined above. Although $b_s$ in **Equation 3** will be systematically smaller than $b_0$ in Mørup's simpler model for strongly interacting systems, namely $T_{MAX} = b_0 T_{dd}$ (where it was estimated to be of the order of 10)[18,65], both parameters have essentially the same meaning, except that the influence of local anisotropy is overlooked in the latter model. Since $T_{dd}$ (dipole-dipole interaction) simply sets the interaction energy scale, $b_S$ gathers everything else affecting the interaction energy, namely, as detailed in Ref. [18], the summation over many pairs of magnetic dipoles (long-range character of the dipolar interaction), the particle size distribution, as well as a factor of the order of unity relating the root-mean-square dipolar interaction energy and the (super)spin-glass ordering temperature. It must be emphasized that the Vogel-Fulcher law [from which Equation 3 is derived] describes the divergence of the relaxation time at a finite temperature $T_{int}$, and therefore Equation 3 is a priori only suitable for the analysis of strongly interacting systems. For non- or weakly-interacting systems, which do not yield such finite-T divergence, the modified Arrhenius expression in Equation 3 should be used instead. In short, the proposed "$T_{MAX} - T_{Bm} = b_s T_{dd}$" relation is one way to introduce the influence of the particle anisotropy in strongly interacting particle systems, where, assuming that $T_{MAX}$ mimics the glass transition $T_g$, it suggests $T_g = T_{Bm} + b_s T_{dd}$, to be compared with Mørup's expression $T_g = b_0 T_{dd}$.

To test our data against Equation 3, Figure 3(c) plots $T_{MAX}$ for both the *bare* and *OA-coated* series as a function of the $T_{Bm}$ measured in the corresponding dilute systems [in turn plotted vs $f_{Co}$ in Figure 3(b) from the data in Figure S2]. It is important to underline the originality of this plot, whose abscissa is not some interaction proxy, as customarily found in the literature,[18,20,22,24,42] but the local anisotropy quantified through $T_{Bm}$. The clearly linear dependence and the fitted slope (close to 1 in both series) are in good agreement with Equation 3, unequivocally evidencing the important role of the single-particle anisotropy in the freezing temperature of strongly interacting assemblies. The intercept of the fits gives the average interaction temperature in the two series ($\overline{T_{int}}$). In the *bare* series, $\overline{T_{int}} = 157$ K, which, considering an average $T_{dd} \approx 40$ K, yields $b_s \approx 3.9$ (the range of individual $b_s$ values in the *bare* series is 2.7 – 5.5). This value is indeed smaller than the $b_0 = T_{MAX}/T_{dd}$ estimates for several concentrated nanoparticle systems reviewed by Mørup in Reference [18], in the range 4 – 8. Moreover, the smaller $b_s \approx 2.1$ of the *OA-coated* series appears consistent with a different qualitative prediction suggested by Mørup's diagram, namely the steeper variation of $T_{MAX}$ vs $T_{dd}$ in systems with strong interactions (providing superspin glass behavior) compared with those with weaker interactions (presenting modified single-particle dynamics). Equation 3, thus, provides a method to estimate the interaction temperature simply from the ZFC curves of isolated and dense systems made of the same nanoparticles. In this regard, note that in dipolarly-interacting systems, as in other systems with similar amounts of ferro- and antiferro-like interactions,[66,67] Curie-Weiss fits cannot determine the magnitude of the interactions.

This leads to the question as to how high the anisotropy barrier must raise (relative to the interparticle interaction) to suppress the collective spin glass-like behavior produced by dipolar interactions. First, note that the naïve notion that dipolar interactions, of the order of magnitude of the dipole-dipole interaction energy ($E_{dd} = k_B T_{dd}$), need to overcome the anisotropy energy barrier ($K_{ef}V$) to produce collective behavior must be dismissed, at least in particle assemblies with a random orientation of the uniaxial anisotropy axes. Note that the first three *bare* nanoparticle systems studied here, with energy ratios $K_{ef}V/E_{dd} \approx 35 \cdot T_{Bm}/T_{dd}$ ranging from $\approx 6.6$ (pure maghemite) to $\approx 118$ ($f_{Co} = 0.19$), show FC-ZFC curves typical of superspin glasses. This is not the case for the *bare* $f_{Co} = 0.23$ sample (with the highest anisotropy), with a FC curve still increasing below $T_{MAX}$ and a broader ZFC maximum. This, together with relaxation features presented below, justifies labelling this sample as "marginal", marking the crossover from collective to single-particle dynamics with increasing $K_{ef}V$. Given the unavoidable distribution in nanoparticle anisotropy barrier, the broader ZFC peak in this sample would reflect the particularly wide distribution of relaxation times arising from a mixture of Néel (higher $K_{ef}V$ particles) and critical (collective) dynamics. Moreover, the single-particle dynamics could operate in few-nanoparticles clusters (dimers, trimers, *etc*) of dipolarly-coupled high anisotropy particles, yielding long relaxation times. It is this heterogeneity in dynamics that defines the mentioned "crossover". Consequently, our experimental data allows us to estimate a crossover energy ratio of $(K_{ef}V/E_{dd})_c \approx 130$. Interestingly, this result is not far from the ratio implied in Mørup's phase diagram [$(K_{ef}V/E_{dd})_c \approx 150$; see Supporting Information], based on the experimental data available at the time, separating (modified) single-particle-like behavior from collective freezing in both Mössbauer and magnetometry experiments.[18]

Unfortunately, the ratio $K_{ef}V/E_{dd}$ is rather inconvenient to experimentally predict the behavior of dense assemblies of nanoparticles. A more suitable parameter can be found rewriting Equation 3, using $T_{Bm} = cT_B$ (with $T_B$ defined as the ZFC peak temperature in the reference system of isolated particles), as

$$\frac{T_{MAX}}{T_B} = c + b_s \frac{T_{dd}}{T_B}, \qquad (4)$$

which suggests the ratio $T_{MAX}/T_B$, more experimentally accessible than $K_{ef}V/E_{dd}$, as a simple parameter to predict the single-particle/collective character of a given nanoparticle system and whether a superspin glass behavior can be expected in relatively dense samples. For example, pressed bare nanoparticles 8 nm in diameter, with $T_{MAX}/T_B \approx 4$, have been previously described as a model superspin glass system.[42] To determine the crossover value we have used the data for $f_{Co} = 0.23$ [namely $c \approx 0.7$, $b_s \approx 5.2$ and $(K_{ef}V/E_{dd})_c \approx 134$, thus $\left(\frac{T_B}{T_{dd}}\right)_c \approx 5.1$], which yields a crossover ratio $\left(\frac{T_{MAX}}{T_B}\right)_c \approx 1.7$. This value is marked by the pink line in **Figure 4**, which gathers the interacting systems studied here (both the *bare* and *OA-coated* series), as well as other dense assemblies characterized by us or other groups.[5,20,23,24,33,40,42,68,69] Although this crossover ratio for collective behavior is far from being a universal [inasmuch as the parameters $c$ and $b_s$ in Equation 4 are not], it can be seen that the threshold value suggested above agrees very well with the reviewed studies (all of them dealing with relatively narrow particle size distributions and reporting $T_{MAX}$, $T_B$ and whether the denser system presents collective or single-particle dynamics). It is particularly noteworthy that another "marginal" sample, reported by Hansen *et al.* in a 5 %vol concentrated FeC ferrofluid) lies also on the same crossover ratio.[5] Moreover, the dashed line in Figure 4 is a fit to Equation 4 of the bare nanoparticle systems studied here with the addition of another disc made of smaller (6.2 nm) maghemite particles.[40] As expected, the resulting value for the fitting parameter $b_s T_{dd}$ is the same (within the error bars) as that obtained for the intercept $b_s T_{dd} = \overline{T_{int}}$ in Figure 3(c) using the equivalent Equation 3. Overall, the data points distribution visually emphasizes the intuitive fact that it is easier to obtain collective behavior in systems comprising nanoparticles with lower anisotropy energy.

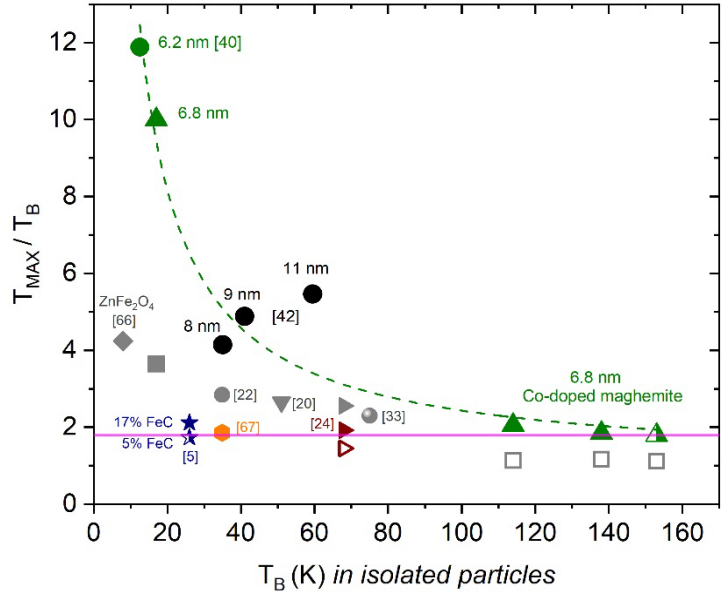

**Figure 4.** Phase diagram of dipolarly interacting nanoparticle systems. The solid symbols correspond to systems with reported collective/superspin glass behavior, whereas hollow symbols are used for systems where the lack of collective behavior has been proven. Half-solid symbols correspond to systems with reported marginal behavior. The pink line, therefore, marks an experimental $T_{MAX}/T_B$ threshold for superspin glass behavior. Larger symbols are used for compacts of essentially *bare* nanoparticles.[22,40,42] Grey symbols are used for pressed OA-coated or dendrimer-coated (down-triangle) nanoparticles.[20,22,24,33,69] Dark blue symbols are for frozen ferrofluids.[5] The nanoparticles are always maghemite[20,22,24,33,40,42,68] unless otherwise noted.[5,69] Importantly, all samples included in the graph were prepared using nanoparticles with a relatively narrow size distribution. The dashed line is a fit of the green datapoints to Equation 4, equivalent to the fit to Equation 3 in Figure 3.

From the crossover ratio $(K_{ef}V/E_{dd})_c \approx 130$ one may also calculate the minimum volume concentration/packing fraction of nanoparticles necessary to reach collective dipolar behavior, $\phi_c$, by simply inserting $E_{dd} = \frac{\mu_0}{4\pi}M_S^2 V \phi_c$,[65] which yields a particle-size independent $\phi_c \approx \frac{0.1\,K}{\mu_0 M_S^2}$. This leads to a realistic $\phi_c \approx 0.58$ for our $f_{Co} = 0.23$ sample. An interesting corollary is that dipolar collective magnetism is not possible (*i.e.*, $\phi_c \gtrsim 0.64$ in random close-packed assemblies) for nanoparticulate materials with $\frac{K}{M_S^2} \gtrsim 8.3$ J kA$^{-2}$m$^{-1}$. This will be the case for hard particles with a modest saturation magnetization, such as ε-Fe$_2$O$_3$ nanoparticles (which yields an impossible $\phi_c \approx 4$ when using the $K$ and $M_S$ values reported in Ref. 69) or CoPt, for which $\phi_c \approx 0.60$ using *bulk* values[55] (higher $K$ and lower $M_S$ values are customarily found in nanoparticles, thus pushing the critical concentration up to unphysical values).

In the discussion of Figure 3 above we claimed the relevance of the particle anisotropy even in systems showing collective dynamics. In this regard, to demonstrate the existence or the lack of such collective behavior, we have performed ZFC memory experiments (encompassing the phenomena of ageing and rejuvenation in spin glass-like systems) probing the system relaxation over several hours. This experiment, together with the previous ZFC/FC magnetization curves, and ac magnetic susceptibility (see SI), cover three of the four "key bulk measurements that determine a magnetic material to be a *canonical* spin glass" (the fourth one being specific heat), according to a recent review by J. A. Mydosh.[71] The ZFC memory effect is a remarkable phenomenon characteristic of the chaotic non-equilibrium (super) spin glass phase.[51,52,72,73] **Figure 5** shows ZFC magnetization curves recorded after cooling without (reference curves) and with (memory curves) a 4 hour halt at $T_{halt} = (2/3)T_{\mathrm{MAX}}$. A dip in the magnetization of the latter curve can be appreciated in the vicinity of the halt. The stop performed during cooling allows the system to "age", slowing down its dynamics towards the particular equilibrium state at the halt temperature; a state that is recovered later upon heating, thus the memory curve magnetization falls below the reference curve. The corresponding difference curves are shown in the lower inset of Figure 5, where Δ*M* is defined as the difference between the memory and reference curves $\left(M_{mem}^{ZFC}(T) - M_{ref}^{ZFC}(T)\right)$ normalized by $M_{ref} = M_{ref}^{ZFC}(T = T_{\mathrm{halt}})$. All samples display visible memory dips; however, their magnitude is strongly dependent on the doping ratio (see upper inset). While the pure maghemite sample exhibits a sharp memory effect (6 %), the Co-doped samples show weaker memory, barely noticeable in the 23% doped sample. This trend can be ascribed to the net increase of the magnetic anisotropy with Co-doping, in turn slowing down the individual nanoparticle relaxation and, thus, the overall collective effect, as have been previously described in both superspin[11,74] and atomic spin glasses.[75] A similar trend (from sharp collective features in the non-doped sample to marginal/anomalous behavior in the 23% doped sample) is observed in the frequency and temperature dependence of the ac susceptibility (see SI).

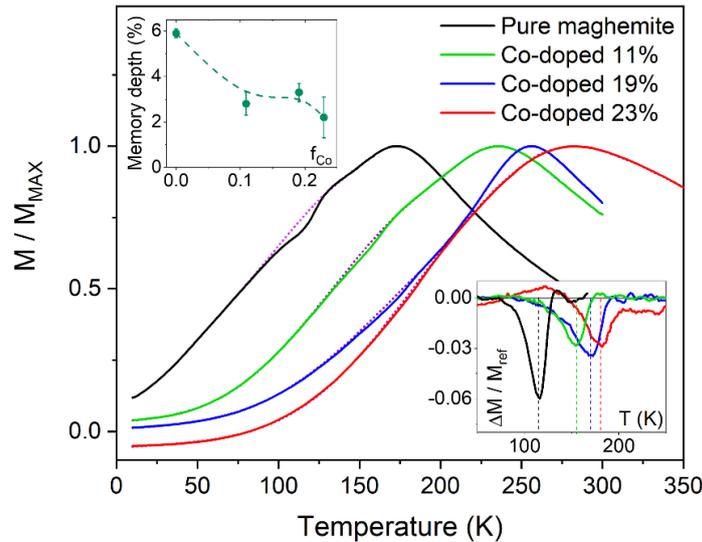

**Figure 5.** ZFC memory experiments. The purple dotted lines are the reference curves measured without a halt. The solid lines are the ZFC magnetization curves measured after 4 hours stops at $T_{halt} = 2T_{MAX}/3$. The lower inset shows the difference curves, $\Delta M(T)/M_{ref}(T = T_{halt})$ ("memory dips"), with the depth plotted in the upper inset. The vertical dashed lines of the lower inset marks $T_{halt}$ for each sample.

## 2.3. Monte Carlo simulations

In addition to the above experiments, the two questions presented in the abstract have also been explored using Monte Carlo simulations of identical spherical magnetic particles with an associated 3D classical macrospin placed at random positions in a cubic lattice (see a detailed description of the model in the SI).[76–83] The Monte Carlo model is designed so that, although it is related to the experiment in terms of nanoparticle size, $M_S$ or $K_{eff}$, it circumvents some of the experimental drawbacks (e.g., variations in the size, $M_S$ or surface disorder between constituents in the series). In this way, we avoid experimental issues that might complicate the interpretation of the data while making the results more general. The Hamiltonian includes only three terms, namely uniaxial anisotropy, dipolar interactions and Zeeman energy. FC and ZFC (with and without a halt to test memory) curves were simulated for ensembles corresponding to the silica-coated (dilute) and bare nanoparticles (60% packing fraction) compacts, where the uniaxial anisotropy was varied systematically. **Figure 6** shows these curves for the end members of the simulated dense series (bare nanoparticles compacts). The qualitative agreement with the experimental data is remarkable, with the highest $K_{ef}$ dense sample showing lower memory effect. Yet more significant, the memory dip progressively fades with increasing anisotropy [see inset in Figure 6(b)], as observed experimentally. At the quantitative front, the dependence of $T_{MAX}$ (from simulations of the nanoparticle compacts) with the mean blocking temperature obtained from the corresponding dilute systems [in analogy to Figure 3(c)] is observed to be linear with a slope very close to unity [see inset in Figure 6(a)], in excellent agreement with the experimental data, thus corroborating the model in Equation 3. Finally, the crossover value $(K_{ef}V/E_{dd})_{c\_MC} \approx 140$ found to transition from collective to individual behavior (corresponding to the anisotropy reducing the memory effect by a factor of three, as suggested by the experimental results) is remarkably similar to our experimental estimate.

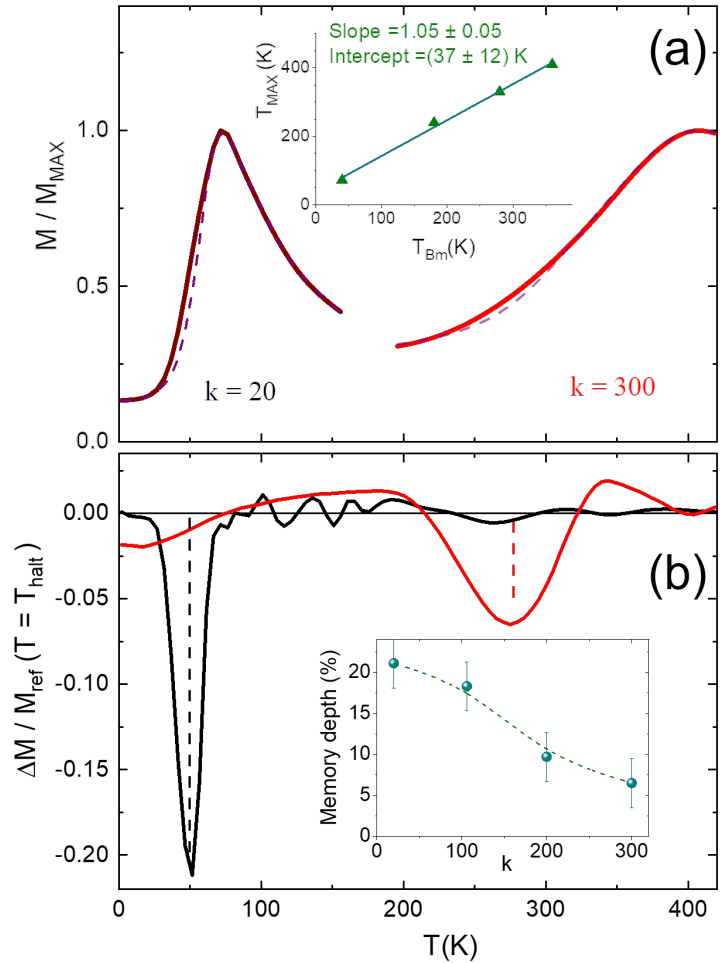

**Figure 6.** (a) FC and ZFC magnetization curves simulated for the end members (lowest and highest $K_{ef}$) of the series of dense systems (same filling factor, varying anisotropy). Solid/dash lines are used to denote ZFC curves obtained without/with a halt, respectively. The inset shows a plot analogous to Figure 3(c), with $T_{Bm}$ calculated from $d(M_{FC}-M_{ZFC})/dT$ as in the experiment. (b) "Memory curves" obtained as the difference between the ZFC curves described above normalized by $M_{ref}(T = T_{halt})$. The vertical lines are the respective halt temperatures. The inset shows the dip depth as a function of the anisotropy constant (defined as $k = K_{ef}V/20k_B$) of the nanoparticles across the simulated series.

## 3. Conclusions

To summarize, we have systematically explored the role of particle anisotropy on the transition from single particle to collective behavior in a series of pellets comprising bare oxide nanoparticles with the same size, size distribution and concentration, but different degrees of Co-doping as a means to control the value of the effective anisotropy ($K_{ef}$). All the compacts made of bare particles display collective glassy behavior (as evident, in particular, from the shape of the FC-ZFC irreversibility and a significant ZFC memory effect), except the sample with the highest nanoparticle anisotropy ($f_{Co}$ = 0.23), which exhibits a "marginal" behavior. The experimental results and the Monte Carlo simulations consistently evidence that: (i) the collective behavior (*e.g.*, freezing temperature, memory effect, ac susceptibility) unquestionably depends on both the dipolar interactions and the nanoparticle anisotropy energy barrier ($K_{ef}V$), and (ii) the minimum $K_{ef}V/E_{dd}$ ratio necessary to suppress the collective behavior in random assemblies of magnetic nanoparticles is of the order of ~100, in good agreement with Mørup's early estimate in his pioneering work on superspin glasses. Our analysis provides (through Equation 3) a method to estimate the interaction temperature simply from the ZFC curves of isolated and dense systems made of the same nanoparticles, and suggests that the ratio $T_{MAX}/T_B$ (easily accessible experimentally) can be used as a predictor of single-particle/collective behavior. The estimated crossover value of ($T_{MAX}/T_B$) ≈ 1.7 separating the two regimes has been successfully checked against a number of previous reports on interacting particle systems with a reasonably narrow particle size distribution (broader distributions will require higher ratios to reach collective behavior). This value leads to the minimum packing fraction of nanoparticles required to yield collective dipolar magnetism, which in fact will never happen in sufficiently hard magnetic materials.

## 4. Materials and Experimental Methods

Monodisperse spherical nanoparticles were synthesized using an optimized thermal decomposition route.[8,40]

Briefly, in a typical synthesis, iron pentacarbonyl and cobalt pentacarbonyl in different proportions ($f_{Co}$) are thermally decomposed in the presence of oleic acid (surfactant) and dioctyl ether (solvent) and subsequently oxidized with trimethylamine N-oxide at high temperature. The nanoparticle size was controlled by changing the amount of oleic acid in the reaction. All the nanoparticles were washed by several cycles of coagulation with acetone, centrifugation, disposal of supernatant solution and re-dispersion in hexane. Then, a fraction of these nanoparticles was washed repeatedly in acetone to remove the oleic acid coating, obtaining what we call here bare particles (≈ 5%w oleic acid), a second fraction was separated to subsequently grow a silica shell following the method described elsewhere,[41] and a third fraction was left as it obtained (oleic acid-coated). All the different types of resulting nanoparticles were dried and the powder was pressed uniaxially under approximately 0.8 GPa to form dense discs.

Transmission electron microscopy (TEM) images have been acquired in a FEI Tecnai G2 F20 microscope operated at 200 kV. The size of the particles was determined by manually measuring the diameters of the particles from the TEM images.

A Quantum Design EverCool MPMS SQUID magnetometer was used for the magnetic characterization of the pellets with the applied field parallel to the disc plane. Zero-field-cooled (ZFC) and field-cooled (FC) magnetization curves were measured using a magnetic field strength of 0.5 mT (5 Oe) and a sweeping rate of 2.5 K/min. In the ZFC protocol the sample was cooled in zero-field from room temperature to 5 K, at which the magnetic field was applied, and the magnetization was recorded upon heating. Then, in the FC protocol, the sample was cooled in constant field and the magnetization recorded again upon heating. ZFC memory experiments were carried out using a similar protocol to that described in Reference [84]. In the "memory curve", the sample was cooled down to 5 K in zero field with a 4 h halt at $T_{halt} = 2T_{MAX}/3$.

**Acknowledgments**

We acknowledge financial support from grant No. MAT2015-65295-R funded by MCIN/AEI/10.13039/501100011033 and by "ERDF A way of making Europe", grant No. PID2019-106229RB-I00 funded by MCIN/AEI/10.13039/501100011033 and the Spanish MEC (through the contract No. BEAGAL18/00095). We also acknowledge funding from UCLM's Plan Propio, the Swedish Research Council (VR), the Universidad Pública de Navarra (grant No. PJUPNA2020) and the Generalitat de Catalunya (grant No. 2017-SGR-292). ICN2 is funded by the CERCA program/Generalitat de Catalunya and supported by SEV-2017-0706 grant funded by MCIN/AEI/10.13039/501100011033.